\documentclass{article}

\usepackage[dblblindworkshop, final]{coml_2025}
\usepackage{algorithm}
\usepackage{algorithmic}
\usepackage{amsmath}
\usepackage{graphicx}
\usepackage{booktabs} 

\workshoptitle{Constrained Optimization for Machine Learning (COML)}

\usepackage[utf8]{inputenc} 
\usepackage[T1]{fontenc}    
\usepackage{hyperref}       
\usepackage{url}            
\usepackage{booktabs}       
\usepackage{amsfonts}       
\usepackage{nicefrac}       
\usepackage{microtype}      
\usepackage{xcolor}         

\title{Causal-Informed Hybrid Online Adaptive Optimization for Ad Load Personalization in Large-Scale Social Networks}

\small{
\author{
  Aakash Mishra\textsuperscript{*} \\
  Meta \\
  \texttt{aakamishra@meta.com} \\
  \vspace{0.2cm}
  \and
  \textbf{Qi Xu}\textsuperscript{*} \\
  Meta \\
  \texttt{xuqi@meta.com} \\
  \vspace{0.2cm}
  \and
  Zhigang Hua \\
  Meta \\
  \texttt{zhua@meta.com} \\
  \vspace{0.2cm}
  \and
  Keyu Nie \\
  Meta \\
  \texttt{keyunie@meta.com} \\
  \vspace{0.2cm}
  \and
  Vishwanath Sangale \\
  Meta \\
  \texttt{vishsangale@meta.com} \\
  \and
  Vishal Vaingankar\\
  Meta \\
  \texttt{vishalv@meta.com} \\
  \and
  Jizhe Zhang \\
  Meta \\
  \texttt{jizhezhang@meta.com} \\
  \and
  Ren Mao\\
  Meta \\
  \texttt{neroam@meta.com} \\
}}

\begin{document}

\maketitle

\begin{abstract}
Personalizing ad load in large-scale social networks requires balancing user experience and conversions under operational constraints. Traditional primal-dual methods enforce constraints reliably but adapt slowly in dynamic environments, while Bayesian Optimization (BO) enables exploration but suffers from slow convergence. We propose a hybrid online adaptive optimization framework \textbf{CTRCBO} ( Cohort-Based Trust Region Contextual Bayesian Optimization), combining primal-dual with BO, enhanced by trust-region updates and Gaussian Process Regression (GPR) surrogates for both objectives and constraints. Our approach leverages a upstream Causal ML model to inform the surrogate, improving decision quality and enabling efficient exploration-exploitation and online tuning. We evaluate our method on a billion-user social network, demonstrating faster convergence, robust constraint satisfaction, and improved personalization metrics, including real-world online AB test results.
\end{abstract}

\renewcommand{\thefootnote}{\fnsymbol{footnote}}
\footnotetext[1]{These authors contributed equally to this work.}

\section{Introduction}

Large-scale social networks serve billions of users daily, requiring personalized decision-making under operational constraints. A prime example is ad load personalization, where showing too few ads underutilizes conversions, while too many degrade engagement and retention. Optimizing ad load thus constitutes a high-dimensional, constrained online optimization problem, where decisions must adapt rapidly to dynamic user behavior while respecting business and user-experience constraints.

Traditional primal-dual optimization methods enforce constraints and provide interpretable dual signals, but they are inherently exploitative, adapting slowly in dynamic, large-scale environments. Bayesian Optimization (BO), on the other hand, enables exploration under uncertainty, but suffers from slow convergence in high-dimensional spaces and is sensitive to surrogate model quality.  

To model the impact of ad load on user engagement and conversions, we leverage an upstream causal ML model~\cite{shi2024ads}, which estimates counterfactual treatment effects under different ad load scenarios. These causal estimates inform Gaussian Process Regression (GPR) surrogates used for both the objective and constraints in our hybrid Primal-Dual + BO framework. The primal-dual solver is enhanced with a trust-region approach for stable updates, while the BO component inherits the surrogate setup, enabling efficient exploration-exploitation and online adaptive tuning. The end to end system illustration is shown in Figure 1.  

We validate our approach on a billion-user social network, showing faster convergence, robust constraint satisfaction, and improved personalization metrics. The method is also evaluated in online production via AB testing, demonstrating its practicality for large-scale deployment.

\paragraph{Contributions:} 
\begin{itemize}
    \item We propose a hybrid Primal-Dual + BO algorithm for real-world online ad load optimization under various business constraints in large-scale social networks.
    \item We show that integrating primal-dual with BO combines their complementary strengths: primal-dual ensures stability and constraint satisfaction, while BO provides adaptive exploration and surrogate-guided tuning.
    \item We adopt the proposed method to a billion-user social network platform, demonstrating improvements in convergence speed, constraint adherence, and personalization performance.
\end{itemize}

\begin{figure}[htbp]
    \centering
    \includegraphics[width=0.8\linewidth]{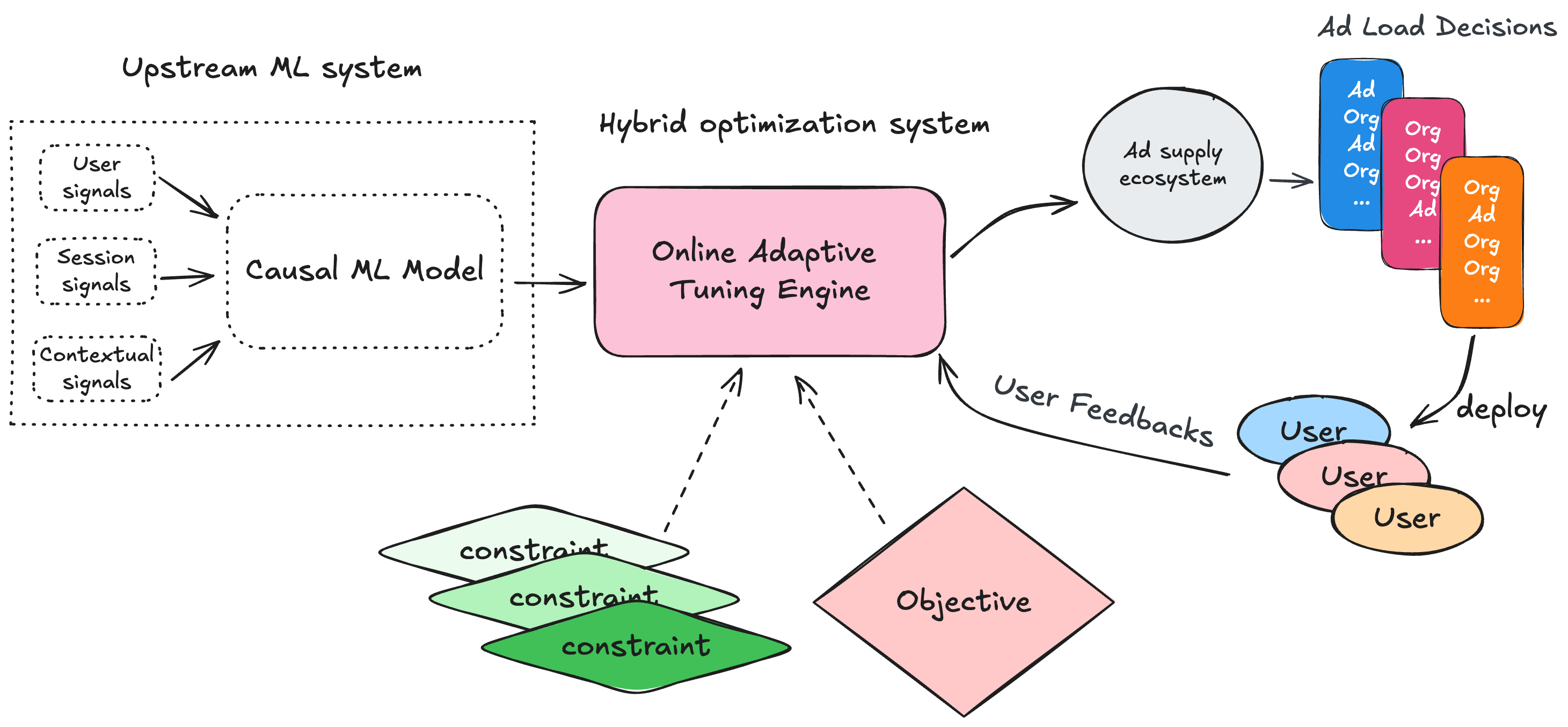}
    \caption{Causal-informed hybrid online optimization system for ad load personalization}
    \label{fig:convergence}
\end{figure}

\section{Related Work}

Constrained optimization is central to ad load tuning in online advertising systems, balancing user experience and conversions. Traditional methods often employ primal-dual algorithms to enforce constraints while optimizing ad placements; for example, Buchbinder et al.~\cite{buchbinder2007online} introduced a primal-dual approach for online ad auctions, and Xu et al.~\cite{xu2024large} extended this to large-scale sponsored search with online adaptation. Recent work highlights the benefit of causal learning for modeling user response to ad exposure, enabling more reliable counterfactual estimates. Goli et al.~\cite{goli2022personalizing} showed that personalized ad load decisions based on causal estimates can optimize conversions, and Sagtani et al.~\cite{sagtani2023ad} used off-policy learning for ad-load balancing. Our work leverages this causal insight as an upstream model to guide GPR surrogates within a hybrid Primal-Dual + BO framework, achieving fast and adaptive constrained optimization at billion-user scale.

\bibliographystyle{plain}

\section{Hybrid Online Adaptive Optimization Framework}


\subsection{Problem Space Formulation}

We implement the hybrid optimization framework by combining the spatial partitioning algorithm provided by multi-objective bayesian optimization (MoRBO) as described in Daulton et al. 2022 ~\cite{daulton2022multi} and the temporal adaptation provided by the primal-dual contextual bayesian optimization (PDCBO) formula as described in Xu et al. 2023 ~\cite{xu2023primal} to achieve high-dimensional contextual multi-objective control. We take advantage of the versatility enabled by trust-region enhancement of contextual methods to enable local modeling across user cohorts while evaluating new policy parameters in parallel for batch contextual optimization. Through our framework, we enable information sharing across A/B testing versions that allow for faster localized policy convergence. 

Consider the following problem formulation: $\boldsymbol{\theta}$ represents a high-dimensional policy vector that determines the placement of ads within the e-commerce system by conditionally weighting session and user-level cohorts which have customized position values. $\boldsymbol{T}$ represents time horizon. $\boldsymbol{N}$ is the total number of constraints. $\boldsymbol{\Theta}$ represents the entire online policy space. $\boldsymbol{z}_t$ represents the time-varying context of the system. $\boldsymbol{R}_T$ is the total cumulative regret at the time horizon. $\boldsymbol{g}_i(\boldsymbol{\theta}, \boldsymbol{z}_t)$ represents the constraint functions dependent on the system context and policy. $\boldsymbol{\gamma}_{i, T}$ is the maximal information gain achievable at a specific observation point. $\boldsymbol{Tr}_j$ represents a trust region with the global parameter space.

\begin{algorithm}
\caption{Cohort-Based Trust Region Contextual Bayesian Optimization (\textbf{CTRCBO})}
\begin{algorithmic}[1]
\STATE \textbf{Input:} User cohorts $\mathcal{C} = \{C_1, \ldots, C_K\}$, trust regions $\{Tr_1, \ldots, Tr_K\}$, 
\STATE \hspace{1.5cm} time horizon $T$, constraint slackness $\epsilon$, learning rate $\eta$
\STATE \textbf{Initialize:} Dual variables $\lambda_1 = \mathbf{0}$, trust region centers for each cohort
\FOR{$t = 1, \ldots, T$}
    \STATE Observe contextual variables $z_t$ (ads impressions shifts, user behavior)
    
    \STATE \textbf{Local GP modeling per cohort-trust region}
    \FOR{cohort $k = 1, \ldots, K$}
        \STATE Fit local GP $f_{k,t}(\theta, z_t)$ for ads score in trust region $Tr_k$
        \STATE Fit constraint GP $c_{k,t}(\theta, z_t)$ for impression budget in $Tr_k$
    \ENDFOR
    
    \STATE \textbf{Primal update with multi-objective acquisition}
    \FOR{cohort $k = 1, \ldots, K$}
        \STATE Compute hypervolume improvement $\text{HVI}_k(\theta, z_t)$ for objectives $(f_k, g_k)$
        \STATE $\theta_{k,t} = \arg\max_{\theta \in Tr_k} \left\{ \text{HVI}_k(\theta, z_t) + \eta \lambda_t^T c_{k,t}(\theta, z_t) \right\}$
    \ENDFOR
    
    \STATE \textbf{Dual update for time-average constraint satisfaction}
    \STATE $\lambda_{t+1} = \left[\lambda_t + \sum_{k=1}^K w_k c_{k,t}(\theta_{k,t}, z_t) + \epsilon \mathbf{e}\right]_+$
    
    \STATE \textbf{Evaluate and update trust regions}
    \STATE Execute policies $\{\theta_{k,t}\}_{k=1}^K$ and observe outcomes
    \FOR{cohort $k = 1, \ldots, K$}
        \STATE Update trust region $Tr_k$ based on success/failure in cohort $k$
        \STATE Update local datasets with new observations
    \ENDFOR
\ENDFOR
\end{algorithmic}
\end{algorithm}

\subsection{Cohort-Based Ads Policy Algorithm}

In order to deliver online policy updates for multi-objective ads supply optimization, we implement the following cohort-based trust region approach that combines contextual Bayesian optimization with time-average constraint handling. Per time step, we observe the contextual information in the system regarding user behavior. We then construct a localized optimization subspace by partitioning the high-dimensional policy space into initial Trust regions ${Tr}_j$ delineated by causal user-cohort sensitivity model that we can explore. Within each subset we fit separate Gaussian Processes (GPs) to model the ads score gains and ad impressions delta. Our local modeling approach provides the same advantages as MORBO and avoids cubic-scaling issues that dominate global GPs for high-dimensional policy constraint spaces. For our policy selection, we execute the primal optimization step that maximizes the hypervolume improvement across the ads score to impressions trade-off while incorporating dual-variables $\lambda_t$ that weight our cumulative constraint violations. Dual-variables are then updated using the PDCBO framework to ensure that we have time average constraint satisfaction weighted by across cohorts by their ads impression volume. After we execute each of the selected policies $\{\theta_{k, t}\}_{k=1}^K$ across all A/B testing arms, we observe the ads score performance and update each cohort's trust region iteratively. 

\subsection{Bounds and Convergence}

Note that the constraint on regret per trust-region ${Tr}_j$ in MORBO is $O(\sqrt{T})$ hypervolume regret and and the constraint on PDCBO $R_T$ is bounded by sub-linear regret $O(\sqrt{\gamma T}\sqrt{T})$ because PDCBO ensures $(1/T) \sum_{t=1}^T g_i(\theta_t, z_t) \leq 0$. Such averaging is necessary in the ads policy tuning space because it is not necessary that a specific policy will be compliant within the given constraints at single given time point (day-over-day). Measurements of ad-load often fluctuate with changing user behaviors and time context $z_k$. In our study we aim to empirically evaluate whether CTRCBO outperforms naive CBO by converging faster. 

\subsection{Gaussian Processor Formulation}

In order to model the ads score and impressions trade-off with  policy parameters, we evaluated our fitting-basis kernel for our GP on observational data at the session level which showed a strong exponential relationship. We define our kernel as $k_{\text{sigmoid}}(x_i, x_j) = \sigma_f^2 \cdot \frac{1}{1 + e^{-(a \cdot x_i^T x_j + b)}} $ where $a$ is the slope parameter, $b$ is the bias parameter and $\sigma_f^2$ represents signal variance. We can tune these parameters to make the GP more / less regularized. For smaller ad surfaces where we encounter more noise, we choose parameters that make the function less susceptible to noise and overfitting. 

\section{Results}

\begin{figure}[htbp]
    \centering
    \includegraphics[width=0.8\linewidth]{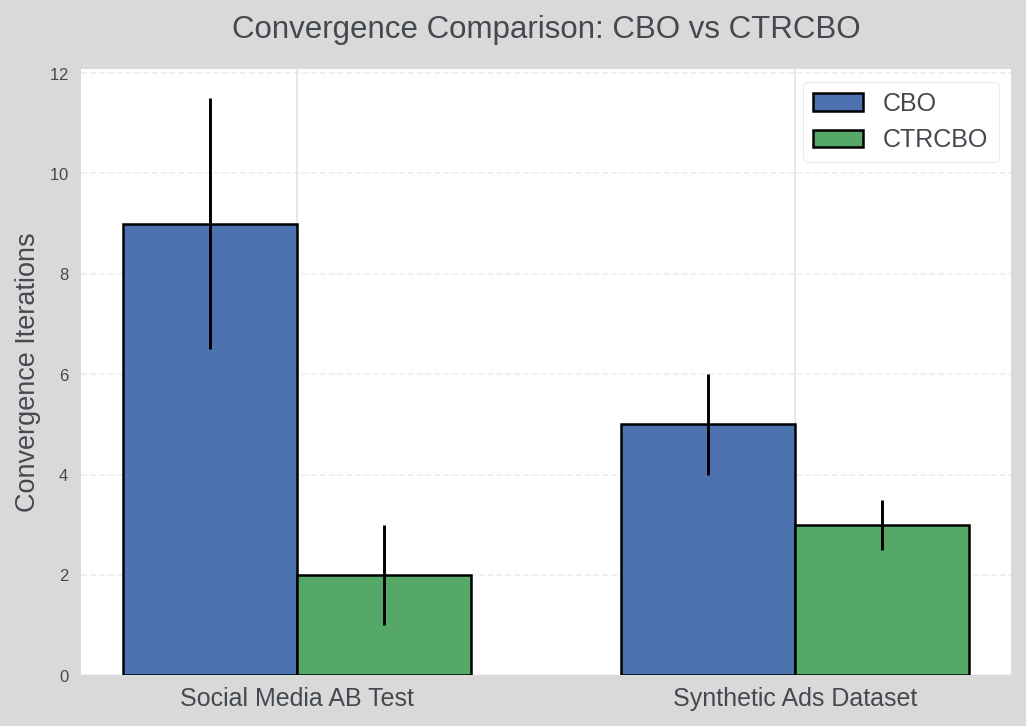}
    \caption{Comparison of convergence iterations between CBO and CTRCBO.}
    \label{fig:convergence}
\end{figure}

We ran an A/B test with several versions across a billion-user scale social media platform to deliver ads based on online policies tuned with CTRCBO. We wanted to see what the convergence of the parameter space would look like under different types of ad impressions constraints and user cohorts. Each experiment had a specific ads score target threshold and ad impressions constraint. If the policy met the ad impressions constraint over a 2-day time interval while keeping the ads score above the target threshold - we would consider objective to be complete. One consideration is that the policy does not have to be globally optimal in order to satisfy the experiment ad impressions constraint and ad score target. However, the policy would have to perform better than the randomized upperbound ad experiment which set a baseline for the naive ad impressions to ad score trade-off. We compared these results with the convergence time for naive CBO. 

\begin{table}[htbp]
    \centering
    \caption{Proxy Model Prediction Comparison on AB Test Results. Target constraint was at least -1\% reduction in ad impressions for a -0.25\% threshold in ads score.}
    \label{tab:example}
    \begin{tabular}{ccc}
        \toprule
        \midrule
         & Ads Impressions  & Ads Score  \\
        \midrule
        Predicted (GP)  & -1.4 & -0.2  \\
        Actual (AB)  & -1.33  & -0.19 \\
        \bottomrule
    \end{tabular}
\end{table}

We also compare these results with the convergence time on our synthetic dataset which was grounded in observational data and used three simulated user sensitivity cohorts with different underlying ads impressions to ads score trade-offs. We wanted to determine how quickly each algorithm would achieve convergence on average to a policy that met the threshold of a 1\% ads score increase for a 1.5\% ads impression increase. 

\subsection{Conclusion}

Overall, we find that the CTRCBO outperforms CBO for online policy tuning based on our casual model cohorts, and our proxy-models show good performance in predicting ads impressions and ads scores conditioning on a given online policy and system context. We demonstrate that the algorithm has faster convergence and robust constraint satisfaction which will help scale ads supply systems.

\bibliography{references}

@misc{shi2024ads,
  author = {Wei Shi and Chen Fu and Qi Xu and Sanjian Chen and Jizhe Zhang and Qinqin Zhu and Zhigang Hua and Shuang Yang},
  title = {Ads Supply Personalization via Doubly Robust Learning},
  year = {2024},
  eprint = {2410.12799},
  archivePrefix = {arXiv},
  primaryClass = {cs.IR},
  url = {https://arxiv.org/abs/2410.12799}
}

@inproceedings{buchbinder2007online,
  title={Online primal-dual algorithms for maximizing ad allocations},
  author={Buchbinder, Niv and Naor, Joseph S and others},
  booktitle={European Symposium on Algorithms},
  year={2007},
  pages={253--265}
}

@article{xu2024large,
  title={Large-scale sponsored search ad allocation with online adaptation},
  author={Xu and others},
  journal={Nature Scientific Reports},
  year={2024},
  volume={14},
  pages={1--12}
}

@article{goli2022personalizing,
  title={Personalizing ad load to optimize subscription and ad revenues},
  author={Goli, A and others},
  journal={Marketing Science},
  year={2022},
  volume={41},
  number={6},
  pages={1012--1029}
}

@article{sagtani2023ad,
  title={Ad-load balancing via off-policy learning in a content marketplace},
  author={Sagtani, Hitesh and others},
  journal={arXiv preprint arXiv:2309.11518},
  year={2023}
}

@inproceedings{daulton2022multi,
  title={Multi-objective bayesian optimization over high-dimensional search spaces},
  author={Daulton, Samuel and Eriksson, David and Balandat, Maximilian and Bakshy, Eytan},
  booktitle={Uncertainty in Artificial Intelligence},
  pages={507--517},
  year={2022},
  organization={PMLR}
}

@inproceedings{xu2023primal,
  title={Primal-dual contextual bayesian optimization for control system online optimization with time-average constraints},
  author={Xu, Wenjie and Jiang, Yuning and Svetozarevic, Bratislav and Jones, Colin N},
  booktitle={2023 62nd IEEE Conference on Decision and Control (CDC)},
  pages={4112--4117},
  year={2023},
  organization={IEEE}
}


\appendix


\end{document}